\documentclass[12pt]{article}

\usepackage{amssymb}
\usepackage{amsmath}
\usepackage{latexsym}
\usepackage{yfonts}

\catcode`@=11
\renewcommand{\section}{\@startsection{section}{1}{0pt}{\medskipamount}
{\medskipamount}{\large\bf}} \numberwithin{equation}{section}
\catcode`@=12
\def\beq{\begin{eqnarray}}    %%%  begequation/eqnarray
\def\eeq{\end{eqnarray}}      %%%  endequation/eqnarray

\def\ln{\,\mbox{ln}\,}                  %%% log
\def\str{\,\mbox{str}\,}                %%% supertrace
\def\sDet{\,\mbox{sDet}\,}              %%% superdeterminant
\def\div{\,\mbox{div}\,}                %%% divergence
\def\pa{\partial}                       %%% partial
\def\={\ =\ }

\begin{document}

\begin{center}

{\Large\bf Closed description of arbitrariness in resolving
quantum master equation}

\vspace{18mm}

{\large Igor A. Batalin$^{(a, b)}\footnote{E-mail:
batalin@lpi.ru}$\; and\;
Peter M. Lavrov$^{(b, c)}\footnote{E-mail:
lavrov@tspu.edu.ru}$
}

\vspace{8mm}

\noindent ${{}^{(a)}}$
{\em P.N. Lebedev Physical Institute,\\
Leninsky Prospect \ 53, 119 991 Moscow, Russia}

\noindent  ${{}^{(b)}}
${\em
Tomsk State Pedagogical University,\\
Kievskaya St.\ 60, 634061 Tomsk, Russia}

\noindent  ${{}^{(c)}}
${\em
National Research Tomsk State  University,\\
Lenin Av.\ 36, 634050 Tomsk, Russia}

\vspace{20mm}

\begin{abstract}
\noindent
In the most general case of the Delta exact operator valued generators
constructed of an arbitrary Fermion  operator,  we present a closed solution
for the transformed master action  in  terms  of the  original master action in
the closed form of the corresponding  path integral.   We show in detail
how that path integral reduces to the known result  in the case of being the
Delta exact generators constructed of an arbitrary Fermion function.
\end{abstract}

\end{center}

\vfill

\\

\\

\setcounter{section}{0}
\renewcommand{\theequation}{\thesection.\arabic{equation}}
\setcounter{equation}{0}

\section{Introduction}

It is recognized commonly that the field-antifield
formalism  in its present form provides  for the most powerful
BRST- inspired methods for covariant  (Lagrangian)  quantization as applied
to complex relativistic  gauge-invariant dynamical systems.

It  is  well  known  that  the gauge invariant status of the general
field-antifield formalism is  completely under control of the quantum master
equation.
The existence of the Fermion nilpotent Delta-operator makes it possible to
expect that the transformations with Delta exact generators do act
transitively
on the set of allowed solutions to the quantum master equation. These
generators have the form of $[ \Delta, F ]$,  where $F$ is a  Fermion operator,
in general.
Usually, one considers  a simple case of being  the $F$ an arbitrary Fermion
function $F( Z )$ rather than an operator \cite{BBD,BLT-BV,BLT-EPJC}.
In the latter case the
corresponding arbitrariness is a set of finite anticanonical master
transformations \cite{BBD,BLT-BV,BLT-EPJC}.  In the simplest case of being  $F( Z )$ only
quadratic in Z, these linear transformations preserve  the antisymplectic
metric, so that we call them  an  antisymplectic rescaling.  We conjecture
that  the field renormalizations  can be included naturally into the group
of antisymplectic rescaling.  In the present article,  our main purpose is
to give a closed description to the arbitrariness in resolving
the quantum master equation in the most general case of being the $F$ an
arbitrary  Fermion  $ZP$ ordered  operator $F( Z, P )$, where $P$ is a canonically
conjugate for $Z$.  Of course,  an explicit solution  is impossible in that
case.  However, by making use of the symbol calculus, together with the
functional methods \cite{Ber1,BSh,Fradkin},  we express the transformed master action in
terms of the original master action in the closed form of the corresponding
path integral.
In principle, the latter path integral  can be calculated, in general ,  in
the form of  quasi-classical   loop  expansion.   On the other hand,  it is
an interesting  question, how the path integral suggested reproduces the
explicit solution for the transformed master action in the previous simple
case of being the $F$
an arbitrary Fermion function $F( Z )$. It appears that in the latter case,
there happens exactly the phenomenon of quantum localization of classical
mechanics \cite{BL}, so that the $P$ integration yields the delta functional
concentrated exactly on the explicit  anticanonically  $F$-transformed  $Z$,
which results in precise
reconstruction to the previous explicit solution.

\section{Antisymplectic rescaling to the quantum master equation}

Let us proceed with the standard quantum master equation
\beq
\label{R1}
\Delta  \exp\left\{ \frac{ i }{ \hbar } W \right\}  =  0,\quad
\varepsilon( \Delta ) = 1, \quad    \Delta^{2} = 0    %   (1)
\eeq
to be resolved for the quantum action $W$,  $\varepsilon( W ) = 0$.
Its natural automorphisms are given by the well-known formula
\cite{BBD,BLT-BV,BLT-EPJC}
\beq
\label{R2}
\exp\left\{ \frac{ i }{ \hbar } W \right\} \; \rightarrow \;
\exp\left\{ \frac{ i }{ \hbar } W' \right\}  =:  \exp\{
[ \Delta, F ] \}  \exp\left\{ \frac{ i }{ \hbar } W  \right\}.  %   (2)
\eeq
The form of a  supercommutaror  of two  Fermion  operators,
with being  at least one of them  nilpotent,  is rather characteristic
for the  unitarrizing
Hamiltonian  in the  generalized  Hamiltonian  formalism
\cite{FVh1,BVh,BFr1,BF2,BF1},  especially,
in the formulation invariant under time reparametrizations \cite{M,M1}.  That  form
is  also  known  to  yield  the  Heisenberg equations of motion  whose right-hand
side is proportional  to  the sum  of the  two  dual   quantum  antibrackets
\cite{BM2,BM5,BL3}  generated,  respectively,  by  each  of
the  two  operators  involved.

It seems natural to conjecture that the renormalization can be
included into the group of antisymplectic rescalings  extracted from
(\ref{R2}) by choosing a quadratic ansatz  for $F$,
\beq
\label{R3}
F =:  \frac{1}{2} Z^{A} F_{AB} Z^{B},  %  (3)
\eeq
\beq
\label{R4}
F_{AB} = {\rm const}( Z ),  \quad \varepsilon( F_{AB} ) = \varepsilon_{A} +
\varepsilon_{B} + 1, % (4)
\eeq
\beq
\label{R5}
F_{AB} = F_{BA} (-1)^{ \varepsilon_{A} \varepsilon_{B} }.  %  {5}
\eeq
Given a constant invertible antisymplectic metric,
\beq
\label{R6}
E^{AB} = {\rm const}(Z), \quad \varepsilon( E^{AB} ) =
\varepsilon_{A} +\varepsilon_{B}
+1,   %(6)
\eeq
\beq
\label{R7}
E^{AB} =
- E^{BA} (-1)^{ ( \varepsilon_{A} + 1 ) (  \varepsilon_{B} + 1 ) },  % (7)
\eeq
the Delta-operator is defined as to the case of trivial measure density,
$\rho = 1$,
\beq
\label{R8}
\Delta  =:  \frac{1}{2} (-1)^{ \varepsilon_{A} } \pa_{A} E^{AB}
\pa_{B}.    %  (8)
\eeq
Then a remarkable formula holds
\beq
\label{R8a}
[ \Delta, F ]  =  ( \Delta F )  -  {\rm ad}( F )
\eeq
with $F$ being an arbitrary Fermion function $F( Z )$ (Section 3),
as well as an
arbitrary $ZP$ ordered  Fermion operator $F( Z, P )$,
where $P$ is canonically conjugate
to $Z$ (Section 4).

In terms of $E^{AB}$ and $F_{AB}$, let us define the
antisymplectic  generator \cite{BL2},
\beq
\label{R9}
G^{A}_{\;\;B} = - G_{B}^{\;\;A} (-1)^{ (\varepsilon_{A} + 1) \varepsilon_{B} },   % (9)
\eeq
where
\beq
\label{R10}
G^{A}_{\;\;B} =: E^{AC} F_{CB},   \quad G_{B}^{\;\;A} =: F_{BC} E^{CA}.     % (10)
\eeq
In terms of the $G_{A}^{\;\;B}$, the right-hand side in (\ref{R2}) rewrites as
\beq
\nonumber
\exp\left\{ \frac{ i }{ \hbar} W' \right\} &=&
\exp\left\{ - \frac{1}{2} G_{A}^{\;\;A} (-1)^{\varepsilon_{A}} -
Z^{A} G_{A}^{\;\;B}\overrightarrow{\partial_B}\right\}
\exp\left\{ \frac{ i }{ \hbar }W \right\} =\\
\label{R11}
&=&
\exp\left\{ - \frac{1}{2} G_{A}^{\;\;A} (-1)^{ \varepsilon_{A} } +
\frac{ i }{ \hbar
} W_{R}  \right\},   % (11)
\eeq
where
\beq
\label{R12}
W_{R} =: W( Z_{R} ),  \quad  Z^{A}_{R} =: Z^{B} ( \exp\{ - G \} )_{B}^{\;\;A} = ( \exp\{
G\} )^{A}_{\;\;B} Z^{B}.   %   (12)
\eeq
Here in (\ref{R12}), the $Z^{A}_{R}$ is just the antisymplectic
rescaling as applied
to $Z^{A}$. Of course, the matrix
\beq
\label{R12a}
S^{A}_{\;\;B}  =:  ( \exp\{ G \} )^{A}_{\;\;B},      %  (2.14)
\eeq
preserves the antisymplectic metric,
\beq
\label{R12b}
S^{A}_{\;\;C} E^{CD} S^{B}_{\;\;D} (-1)^{ \varepsilon_{D} ( \varepsilon_{B} + 1 ) }
=  E^{AB}.     % (2.15)
\eeq

\section{The general case of an arbitrary Fermion function $F(Z)$}

Now, let us describe in short the general case of arbitrary Fermion function
$F( Z )$ in formula  (\ref{R2}).
Then the formula (\ref{R11}) generalizes as
\beq
\label{R13}
\exp\left\{ \frac{ i }{ \hbar } W' \right\}  =
J^{ 1/2 } \exp\left\{ \frac{ i }{ \hbar } W_{R}\right\},  % (13)
\eeq
where
\beq
\label{R14}
W_{R}  =  W( Z_{R} ),    \quad   Z^{A}_{R}  =  \exp\{ - {\rm ad}( F ) \}  Z^{A},  % (14)
\eeq
\beq
\label{R15}
J =:  \sDet [  (  Z^{A}_{R}\; \overleftarrow{\partial_B}  )  ],  \quad
J^{ 1/2 }  =\exp\{ ( E( - {\rm ad}( F ) ) \Delta F ) \},   %  (15)
\eeq
\beq
\label{R16}
E( X ) =:  \int_{0}^{1} dt  \exp\{ t X \}  =  \frac{ \exp\{ X \} - 1 }{ X }. % (16)
\eeq

\section{The most general case of an
arbitrary Fermion operator $F(Z,P)$}

Finally, let us mention in short the case of being the $F$ an operator,
\beq
\label{R17}
F   =  F ( Z, P ),  \quad [ Z^{A}, P_{B} ]  =  i  \hbar \;\!\delta^{A}_{B} \;\! 1,  %(17)
\eeq
with $P_{A}$ being momenta canonically conjugate to $Z^{A}$,
\beq
\label{R18}
P_{A}  =:  - i \hbar \overrightarrow{\partial_A} (-1)^{ \varepsilon_{A} }.   % (18)
\eeq
In terms of the symbol chosen, say $ZP$ symbol, the formula (\ref{R2}) rewrites as
\beq
\label{R19}
\exp\left\{ \frac{ i }{ \hbar} W' ( Z ) \right\}  =
\left(  \exp_{*}\{ [ {\rm symbol} \Delta,  {\rm symbol} F ]_{*} \}  *
\exp\left\{ \frac{ i }{ \hbar
} W( Z ) \right\}  \right)\!\Big|_{ {\rm symbol}\;\! P = 0 }  ,   %  (19)
\eeq
where $*$ means the symbol multiplication,
\beq
\label{R20}
{\rm operator} A \;\leftrightarrow\; {\rm symbol} A,\quad
{\rm operator} A \; {\rm operator} B \;\leftrightarrow\;
{\rm symbol} A  * {\rm symbol} B,
%(20)
\eeq
$[ \;, ]_{*}$ means the respective symbol supercommutator, and $\exp_{*}$ means
the symbol exponential.
Given the operator $F$ in $ZP$ normal form, let us denote its $ZP$ symbol in short
as $F( Z, P )$,
while the respective symbol multiplication is given by
\beq
* =:   \exp\left\{  - i  \hbar  \frac{\overleftarrow{\partial } }{ \partial P_{A} }
 (-1)^{ \varepsilon_{A} }
\frac{\overrightarrow{ \partial } }{ \partial  Z^{A} }   \right\}. %          (4.5)
\eeq
Then, by proceeding with the symbol representation (\ref{R19}), and using the
standard functional methods \cite{Ber1,BSh,Fradkin}, one can derive
the following path integral solution
\beq
\label{R21}
\exp\left\{ \frac{ i }{ \hbar } W '(Z) \right\}   =
\left\langle \exp\left\{  - \frac{ i }{ \hbar }
\int_{0}^{1}  dt  H( Z( t ), P( t - 0 ) )  +
\frac{ i }{ \hbar } W( Z ( 0 ) )  \right\} \right\rangle,  %   (4.5)
\eeq
\beq
\label{R21a}
H( Z, P )  =:
(  F( Z, P ), Z^{A}  )  P_{A}  (-1)^{ \varepsilon_{A} }  -
\frac{ \hbar }{ i } ( \Delta F( Z, P ) ),   %   (4.6)
\eeq
where the functional average is defined as
\beq
\label{R22}
\langle( ... ) \rangle   =:   \frac{ \int  \mathcal{D} Z  \mathcal{D} P   ( ... )
\exp\left\{ \frac{ i }{ \hbar }  \int _{0}^{1}  dt  P_{A}  \dot{Z}^{A} \right\}}
{ \int  \mathcal{D} Z  \mathcal{D} P
\exp\left\{ \frac{ i }{ \hbar }  \int_{0}^{1}  dt  P_{A}  \dot{Z}^{A} \right\}}, %(22)
\eeq
where the integration trajectory $Z^{A}( t )$ is restricted to satisfy the
condition
\beq
\label{R23}
Z^{A}( t + 0  =  1 )  =  Z^{A} .   % (23)
\eeq
One can take the latter condition into account explicitly by introducing the
well defined representation
\beq
\label{R24}
Z^{A}( t ) =:  Z^{A}  -  \int _{ t + 0 }^{1}  dt'  V^{A}( t' ).      % (24)
\eeq
Then one can change for the integration over unrestricted
velocities $V^{A}(t)$, ${\cal D}Z\rightarrow {\cal D}V$.

Now, let us return temporary to the case of $P$-independent  $F$,  $F = F( Z )$.
Then the $P$ integration in (\ref{R21}) yields the delta functional
\beq
\label{R25}
\delta [ \dot{Z}^{A} - ( F, Z^{A} ) ],     %  (25)
\eeq
so that
\beq
\label{R26}
Z^{A}( t )  = \exp\{ - ( 1 - t ) {\rm ad} ( F ) \} Z^{A},  \quad
Z^{A}( 0 )  =  \exp\{ - {\rm ad}( F ) \} Z^{A}.    % (26)
\eeq
Let us represent the Jacobian of the delta functional (\ref{R25}) via the
unrestricted velocity $V^{A}( t )$,
\beq
\label{R27}
\sDet [ \delta^{A}_{B} \;\!\delta ( t - t' )  +  ( F, Z^{A} )\;\!
\overleftarrow{\partial_B}(Z(t)) \;\theta( t' - t - 0 ) ].   %    (27)
\eeq
By expanding the logarithm of the Jacobian (\ref{R27}) in powers of the second
term, one can easily
see that all orders are zero due to the specific products of the
theta functions. For the first order we have
\beq
\label{R28}
- \int_{0}^{1} dt ( \Delta F)( Z( t ) ) \theta( - 0 )  =  0.    %  (28)
\eeq
For the second order we get
\beq
\label{R29}
\int_{0}^{1} dt  \int_{0}^{1} dt' (  ( F, Z^{A} )
\overleftarrow{\partial_B} ( Z ( t ) )  )
(  ( F, Z^{B} ) \overleftarrow{\partial_A} ( Z( t' ) )  )  (-1)^{\varepsilon_{A}}
\theta( t' - t - 0 )\; \theta( t - t' - 0 )  =  0,    %   (29)
\eeq
and so on (for closed derivation see Appendix A). Thus,  the Jacobian  (\ref{R27})
equals to one. Then, by substituting
the solution (\ref{R26}), we arrive at  the formula (\ref{R13}).
That is a particular case of the phenomenon of quantum localization of
classical mechanics \cite{BL}.

In a purely formal sense, the path integral (\ref{R21}) resolves the Schr\"{o}dinger
equation
\beq
\label{R30}
i  \hbar \partial_{ t } \Psi ( t, Z )  =  H( Z, P )  \Psi( t, Z ),   % (4.15)
\eeq
with $H( Z, P )$ being the operator valued Hamiltonian ( see (\ref{R18}) for momenta
$P_{A}$ ),
\beq
\label{R31}
H( Z, P )  =  ( i \hbar )^{-1} [  F( Z, P ),  \frac{1}{2} P_{A} E^{AB}
P_{B} (-1)^{ \varepsilon_{B} } ],   %   (4.16)
\eeq
where $F( Z, P )$ is assumed $ZP$ ordered,
\beq
\label{R32}
F( Z, P )   =:   F( Z, Y )  \exp\left\{  \frac{\overleftarrow{\partial } }{ \partial
Y_{A} }  P_{A}  \right\} \Big|_{ Y  =  0 }.  %   (4.17)
\eeq
Then  we have
\beq
\label{R33}
\Psi( 0, Z )  =  \exp\left\{ \frac{ i }{ \hbar } W( Z ) \right\},\quad
\Psi( 1, Z )  =\exp\left\{ \frac{ i }{ \hbar } W'( Z ) \right\}.  %   (4.18)
\eeq
Thus,  we  see that in the case of being the  $F(Z,P)$  just an operator
valued quantity,  the arbitrariness  in resolving the quantum master
equation can only
be  described comprehensively by applying  the quantum-mechanical  treatment
in its  precise  form.

Notice that the path integral solution (\ref{R21}) rewrites  naturally into its
variation-derivative form,
\beq
\label{R34}
&&\exp\left\{ \frac{ i }{ \hbar } W'(Z) \right\}  =
\exp\left\{  - \frac{ i }{ \hbar }  \int_{0}^{1}
dt  H\left(  i  \hbar \frac{ \delta }{ \delta J( t ) },   i  \hbar \frac{ \delta
}{ \delta K( t - 0 ) }  \right)  \right\}\times\\
\nonumber
&&
\exp\left\{  - \frac{ i }{ \hbar }  \int_{0}^{1}  dt   J_{A}( t ) \! \left(  Z^{A}   -
\int_{ t }^{1}   dt'   K^{A}( t' )  (-1)^{ \varepsilon_{A} }  \right)  +
 \frac{ i }{ \hbar }  W\!\!\left(  Z  -  \int_{0}^{1}  dt'   K( t' )  (-1)^{
\varepsilon } \! \right) \! \right\} \!\!\Big|_{  J  =  0,  K  =  0 }.      %  (4.20)
\eeq
One can always return back to (\ref{R21}) by inserting the factor
\beq
\label{R35}
1  =  {\rm const}\int \mathcal{ D }V  \int \mathcal{ D }P  \exp\left\{  \frac{ i }{ \hbar }
\int_{0}^{1}  dt  P_{A}  \big( V^{A} - K^{A} (-1)^{ \varepsilon_{A} } \big)  \right\},  %(4.21)
\eeq
to the right of the first exponential in the right-hand side in (\ref{R34}).
Here in (\ref{R35}),  {\rm const}  is a normalization constant.

In the most general case of a non-constant antisymplectic  metric $E^{AB}( Z )$
 and a measure density $\rho( Z )$,
where the Delta operator (\ref{R8}) becomes
\beq
\label{R36}
\Delta  =:  \frac{1}{2}  (-1)^{ \varepsilon_{A} }
\rho^{-1} \partial_{A} \rho E^{AB} \partial_{B},  %    (4.22)
\eeq
the $ZP$ symbol (\ref{R21a}) generalizes as
\beq
\label{R37}
i\hbar\;\! H( Z, P )  &=:&\Pi(Z,{\tilde P})F(Z,P)+F(Z,P)\Pi({\tilde
Z},P), \eeq \beq \label{R37a} \Pi(Z,P)= \frac{1}{2} \big(  E^{AB}( Z
) P_{B} P_A   - i  \hbar ( \div E )^{B}( Z )P_{B}\big)  (-1)^{
\varepsilon_{B} }, \eeq where we have denoted \beq \label{R38}
\tilde{ P }_{A}  =: P_{A}  -  i \hbar \frac{
\overrightarrow{\partial } }{ \partial Z^{A} }
(-1)^{ \varepsilon_{A} },     % (4.24)
\quad
\tilde{ Z }^{A}  =
:  Z^{A}  -  i  \hbar  \frac{ \overleftarrow{ \partial } }{ \partial P_{A} }
(-1)^{ \varepsilon_{A} },   % (4.25)
\eeq
\beq
\label{R40}
(\div E )^{B}( Z )  =:
\rho^{-1}( Z ) \left(  \frac{ \partial }{ \partial Z^{A} }
\rho( Z ) E^{AB}( Z )  \right) (-1)^{ \varepsilon_{A} }.    % (4.26)
\eeq
In its turn, the operator valued Hamiltonian (\ref{R31}) generalizes as
\beq
\label{R40}
H( Z, P )  =  ( i \hbar )^{-1}
[  F( Z, P ), \Pi(Z,P)] .   %   (4.27)
\eeq

In its general features, the above consideration was addressed
to the case of the Delta operator (\ref{R36}) as assumed  to be a nilpotent one.
However,
there exists a bit modified  version  as  to the Delta  operator
\cite{BB} (see also the references therein).
One cancels  the nilpotency assumption for the original Delta operator (\ref{R36}),
and  then defines  a  new nilpotent operator by adding  a
Fermion  function to the (\ref{R36}),  so that  the new Fermion function
is determined just via  the nilpotency condition for the new  Delta operator.
In this way,  the measure density becomes independent of  the antisymplectic metric.
By proceeding with the
new  Delta operator,  one can  apply  the above consideration in a quite similar way.
As a result,  there will  be  no  modifications,  being  the $F( Z )$  a
function.  In the case of being the $F( Z, P )$ an actual  operator,
a simple new term  should  be  added  in the right-hand side in (\ref{R37a}),
that is  $ ( i \hbar )^{2} \;\!\nu( Z )$,
with $\nu( Z )$ being just  the new Fermion function added to the (\ref{R36}).

In the case of a non-constant $\rho( Z )$, as the scalar product is defined
with respect to the invariant integration measure $d \mu( Z )  =:  \rho( Z)
dZ$,  to make the operator (\ref{R18}) Hermitian, the latter  should be transformed
:
\beq
\label{R40a}
P_{A} \;\rightarrow \; \rho^{ - 1/2 }  P_{A}  \rho^{ 1/2 }  =
P_{A}  -  \frac{ i\hbar }{2}  ( \ln  \rho ) \overleftarrow{\partial_A},
\eeq
which results in our having chosen  the same shift as to the  $P$ -argument in every $ZP$ symbol.  The
latter common  shift  can be canceled by  the opposite shift for $P$ in the
kinetic exponential in the *nominator* in (\ref{R22}).  As  a  result, one
acquires the factor
\beq
\label{R40b}
\frac{ \rho^{1/2}( Z( 0 ) ) }{ \rho^{1/2}( Z ) }
\eeq
in
front of the exponential inside the average in the right-hand side in (\ref{R21}).
Thereby,  the equation  (\ref{R21})  takes the form of a transformation law as
formulated for  the  semi-density $ \exp\{  \frac{ i }{ \hbar } W \}\rho^{1/2}$.
Notice  that  for  a  symbol  $F( Z, P )$,  the  above  shift  (\ref{R40a})  does
*not*  coincide  with the  precise form of  the  symbol  transformation
\beq
\label{R40c}
F( Z, P )  \;\rightarrow\; \rho^{ - 1/2 }( Z ) *  F( Z, P ) * \rho^{1/2 }( Z )  =
F(  Z,  P  -  \frac{ i \hbar }{ 2 } ( \ln \rho ) \overleftarrow{\partial_A}
( \tilde{ Z } )  ),     %   (4.30)
\eeq with the $\tilde{ Z }^{A}$  being  given by the second in
(\ref{R38}).  In contrast to the latter formula (\ref{R40c}),  the
second term in (\ref{R40a}) is taken at the original argument
$Z^{A}$,  *not* at the  $\tilde{Z}^{A}$.  However,  in  the
integrand  of  the *regularized*  functional  integral  in
(\ref{R21}),  just the above  shift  (\ref{R40a}) in the symbol
$H(Z, P )$ results,  when having  been canceled via the opposite
shift in  $P_{A}$,  in  appearance  of  the correct  factor
(\ref{R40b}) as  having  it  come from  the kinetic exponential  in
the functional integrand in the nominator in (\ref{R22}).  Notice
also that  the components  of the second  argument of  $F$  in
(\ref{R40c})  do commute among  themselves.

Let us notice by the way that  the above consideration extends
naturally as to the case of  the  $Sp(2)$  symmetric  quantum master
equation \cite{BLT,Hull,BMS,BM,BBL},
\beq
\label{R41}
\Delta_{+}^{a}   \exp\left\{ \frac{ i }{ \hbar }  W \right\}   =   0  ,\quad
\Delta_{+}^{a}  \Delta_{+}^{b}  +  ( a \leftrightarrow b )   =   0,   %  (4.27)
\eeq where  $\Delta_{\pm}^{a}$  is  a pair of the $Sp(2)$-vector
valued Delta operator  together with  its  transposed,
\beq
\label{R42} \Delta_{\pm} ^{a}   =:   \Delta^{a}   \pm  \frac{ i }{
\hbar }  {\cal V}^{
a },\quad \varepsilon ( \Delta^{a} )  =  \varepsilon( {\cal V}^{a} )  =  1, % (4,28)
\eeq
\beq
\label{R42a}
\Delta^{a}  =:  \frac{1}{2} (-1)^{ \varepsilon_{A} } \rho^{-1} \partial_{A}
\rho  E^{ a AB } \partial_{B}  =
\frac{1}{2}  \big( \;\! (-1)^{\varepsilon_{ A } } E^{ a AB } \partial_{B}
\partial_{A}  +  (\div E^{ a })^B \partial_{B} \;\!\big),
\eeq
\beq
\label{R42b}
(\div E^{ a })^B  =:  (-1)^{ \varepsilon_{A} } \rho^{-1} ( \partial_{A} \rho
E^{ a AB } ),
\eeq
\beq
\label{R42c}
\mathcal{
V }^{a}  =:  V^{a}  +  \frac{1}{2}  \div V^{a},\quad
V^{a}   =:   V^{ a A } \partial_{A},\quad  \div V^a =:\rho^{-1}  \partial_{A} (\rho V^{aA}),
\eeq
with  $V^{a}$  being  the  special vector field.  A  counterpart to
the formula (\ref{R2}) has the form \cite{BLT1}
\beq
\label{R43}
\exp\left\{  \frac{ i }{ \hbar }  W  \right\}\;   \rightarrow \;
\exp\left\{  \frac{ i }{ \hbar }W'  \right\} =:
\exp\left\{i  \hbar \;\!  \frac{1} {2} \varepsilon_{ab} [ \Delta_{+}^{b}, [
\Delta_{+}^{a},  B ] ]  \right\}   \exp\left\{  \frac{ i }{ \hbar }  W  \right\}, \quad
\varepsilon(B)=0. %(4.28)
\eeq
By making  use  of the methods quite similar to the above, one can
derive in a simple way a natural counterpart to the formulae (\ref{R21}),
(\ref{R37}).

Of course, the general formula (\ref{R21}) remains valid,  while the formula
(\ref{R37}) generalizes as to take the form
\beq
\label{R43a}
i  \hbar \;\!H(Z,P)=\Pi^b(Z,{\tilde P})F_b(Z,P)+F_b(Z,P)\Pi^b({\tilde Z},P),
\eeq
where
\beq
\label{R43b}
i  \hbar \;\!F_b(Z,P)=\Pi^a(Z,{\tilde P})B(Z,P)\frac{1}{2}\varepsilon_{ab}-
B(Z,P)\frac{1}{2}\varepsilon_{ab}\Pi^a({\tilde Z},P),
\eeq
and
\beq
\nonumber
\Pi^a(Z,P)&=& \frac{1}{2} \big[ \big(  E^{ a AB }( Z )P_{B}
P_{A}  +
(  2 V^{ a B }( Z )  -  i  \hbar\;\! (\div E^{a} )^{B}( Z )  ) P _{B}
\big)  (-1)^{ \varepsilon_{B} }-\\
\label{R43c}
&&\qquad - i \hbar\;\!  \div V^{a}(Z) \big].
\eeq

%\noindent
%{\bf Note added in proof}.
In  the main body of the present  paper,  we have used the normal
$ZP$-symbol,  which is the simplest one technically.  In principle, one could
use another type of symbols, say, the  Weyl  symmetric  symbol.  At  least,  in  the
case of  being  the  generator an arbitrary   Fermion  function $F( Z )$,  it
can  be shown that with  the  use of a new symbol  one reproduces the same formula  (\ref{R15})
in  new co-ordinates.

\section*{Acknowledgments}
\noindent
I. A. Batalin would like  to thank Klaus Bering of Masaryk
University for interesting discussions.
The work of I. A. Batalin is
supported in part by the RFBR grants 14-01-00489 and 14-02-01171.
The work of P. M. Lavrov is supported in part by the RFBR grants 15-02-03594 and
16-52-12012-NNIO and Deutsche Forschungsgemeinschaft (DFG) grant LE 838/12-2.

\appendix
\section*{Appendix A. Closed derivation to the Jacobian (\ref{R27})}
%\section{}
\setcounter{section}{1}
\renewcommand{\theequation}{\thesection.\arabic{equation}}
\setcounter{equation}{0}

Here we present in short a  closed derivation to the Jacobian (\ref{R27}).  Let
us denote
\beq
\label{A.1}
X^{A}_{\;\;B}( t )  =:  ( F , Z^{A} )\;\overleftarrow{\partial_B} ( Z( t ) ). % (A.1)
\eeq
Then, in short matrix notations, logarithm of the Jacobian (\ref{R27}) reads
\beq
\label{A.2}
\ln J   =:  \int _{0}^{1} d\lambda  \int_{0}^{1}  dt  \int_{0}^{1}  dt'
\str(G( t, t' ; \lambda )  X ( t' ) ) \theta( t  -t' - 0 ),     %  (A.2)
\eeq
where the Green's function $G( t, t' ; \lambda)$  is defined by the integral
equation
\beq
\label{A.3}
\int_{0}^{1}  dt'  [  \delta( t - t' )  1  +  \lambda  X( t ) \theta( t' - t
-  0 )  ] G( t', t''; \lambda )  =  \delta(  t  -  t'' )  1 .     %   (A.3)
\eeq
Let us denote
\beq
\label{A.4}
\Gamma( t, t'' ; \lambda )  =:  \int_{ t + 0 }^{1}  dt'  G( t', t''; \lambda),  %    (A.4)
\eeq
then the equation (\ref{A.3}) rewrites in its differential form
\beq
\label{A.5}
[  - \partial_{ t}  + \lambda  X( t )  ]  \Gamma( t, t'' ;  \lambda )  =
\delta(  t  -  t'' ),   \quad    \Gamma(  t + 0  =  1,  t'';  \lambda )  =  0,  %  (A.5)
\eeq
which resolves in the form
\beq
\label{A.6}
\Gamma( t, t'' ; \lambda )  =  \theta( t'' -  t  -  0 )  U( t, t''; \lambda),   %    (A.6)
\eeq
where the holonomy matrix $U$ is defined by the equation
\beq
\label{A.7}
[ - \partial_{ t }  + \lambda  X( t )  ] U( t, t'';  \lambda ) = 0,\quad
U(t  =  t'',  t'';  \lambda )  = 1.  %    (A.7)
\eeq
At  $Z^{A}( t )  =  Z^{A}( t; \lambda )$ in (\ref{A.1}), the
latter Cauchy problem (\ref{A.7}) resolves in the form \beq
\label{A.7a} U( t, t'';  \lambda )  =:  U( t; \lambda ) U^{-1}( t'';
\lambda ), \quad  U( t; \lambda )  =:  Z( t; \lambda )  \otimes
\frac{\overleftarrow{\partial } }{
\partial  Z  },  %    (A.8)
\eeq
where  $Z^{A}( t; \lambda )$  is given by the first in  (\ref{R26})  with the  $F$
being  $\lambda$-rescaled  as $F \rightarrow \lambda  F$ .

It follows from (A.4), (A.6) that
\beq
\label{A.8}
G(t, t' ;  \lambda ) =  -  \partial_{ t } \Gamma( t, t' ;  \lambda )  =
 \delta( t - t' ) 1  -  \theta( t' -  t  -  0 ) \lambda  X( t ) U( t, t' ;
\lambda ).    %   (A.8)
\eeq
By inserting (\ref{A.8}) into (\ref{A.2}) , we arrive at
\beq
\nonumber
\!\!\ln J \!&=& \!\int_{0}^{1}  d\lambda  \int _{0}^{1}  dt  \int_{0}^{1}  dt'
\str \big( \;\! (  \delta ( t  -  t' ) 1
-\theta ( t'  -  t -  0 ) \lambda  X( t )
U( t, t';  \lambda  ))X( t' )\;\!\big)\times\\
\label{A.9}
&&\qquad\qquad\qquad\qquad\qquad\times
\theta( t  -  t'  - 0  )   =   0.  % (A.9)
\eeq
Thus, we have confirmed via closed derivation  that  $J  =  1$.
Notice also that (\ref{A.2}) rewrites directly in terms of (\ref{A.4}) as
\beq
\label{A.10}
\ln J  =  \int_{0}^{1}  d\lambda  \int_{0}^{1}  dt'  \str( \Gamma( t', t' ;
\lambda )  X( t' ) ).   %   (A.10)
\eeq
On the other hand, it follows from (\ref{A.6}) together with the second in (\ref{A.7})
that
\beq
\label{A.11}
\Gamma ( t', t' ; \lambda )  =  \theta( - 0 )  1  =  0,    %  (A.11)
\eeq
which, when being substituted into (\ref{A.10}), confirms (\ref{A.9}), even in a simpler way.

It is worth to mention that the result (\ref{A.9}) seems a bit paradoxical,  as
the  Delta  functional (\ref{R25}) is concentrated on the solution (\ref{R26}) being
an anti-canonical  transformation as applied to $Z^{A}$.  On the other hand, at
the level of the  $Z^{A}$ -space,  an anti-canonical  transformation is known
to yield a nontrivial Jacobian, in general (see the formula (\ref{R15})).  It appears, however,
that at the level  of  the functional space of trajectories $Z^{A}( t )$,  the
corresponding functional  Jacobian  is  trivial,  just  due  to  the presence of  the
theta-functions  regularized  in  accordance  with  the  $ZP$ normal  ordering
chosen.

Finally,  let  us  consider  in  more  details  the relation between  the
functional  Jacobian  (\ref{A.10})  and  the  finite-dimensional  Jacobian  in
the  second  in (\ref{R15}).
Due  to  (\ref{A.11}),  we  have  for  (\ref{A.10}),  now re-denoted  as  $(J')^{-1}$  for
further  convenience,
\beq
\label{A.13}
\ln J'  = - \theta( - 0 )  \int_{0}^{1}  d\lambda  \int _{0}^{1}  dt'  \str( X(
t' ) ),
\eeq
where  $X( t )$  is  defined  in  (\ref{A.1})  with  the  $Z^{A}( t )$  given  by  the
first in  (\ref{R26}).  Thus, the $\lambda$ - integral  is  trivial,
%\beq
%\label{A.14}
%\int_{0}^{1}  d\lambda   =   1,
%\eeq
and  we  rewrite (\ref{A.13})  in  the  form
\beq
\label{A.15} (
J')^{1/2}   =  \exp\{  \theta( - 0 ) ( E( - {\rm ad}( F ) ) \Delta
F ) \}.
\eeq
By  comparing  the  latter  to  the  second  in
(\ref{R15}),  we conclude  that \beq \label{A.16} J'  =  ( J )^{
\theta( - 0 ) }  =  1. \eeq It  is  just  the  relation  that  shows
us  a  miraculous phenomenon of the functional  Jacobian  $J'$,
inverse to (\ref{A.10}), as  having  gulped  the  finite-dimensional  Jacobian  $J$,
the second in (\ref{R15}). The  inverse
functional Jacobian   $J'$  just  comes  to  stand  in  front of the
delta  functional   (\ref{A.10}),  and thus  the  $J'$ is  a natural
candidate  to be  compared  to  $J$  in  (\ref{R15}).

\end{document}